\documentclass[pra,nofootinbib,floats,aps,twocolumn,tightenlines,superscriptaddress]{revtex4-1}

\usepackage{graphicx}
\usepackage{bbold}
\usepackage{mathtools}
\usepackage{hyperref}
\usepackage{commath}
\usepackage{bm}
\usepackage{amsfonts,amsmath,amssymb,amsthm}
\usepackage[usenames,dvipsnames]{xcolor}
\usepackage{subfigure}
\usepackage{dsfont}

\DeclareMathOperator{\Tr}{Tr}
\DeclareMathOperator{\Realpart}{Re}

\renewcommand*\d[2][]{%
	\mathrm{d}%
	\ifx\relax#1\relax\else
	\rule{-0.02em}{1.5ex}^{#1}\rule{0.08em}{0ex}\!
	\fi
	#2\,
}
\newcommand{\ket}[1]{| {#1} \rangle}
\newcommand{\bra}[1]{\langle {#1} |}

\newcommand{\ii}{\mathrm{i}}

\renewcommand{\a}[1]{\hat{a}_{\bm{#1}}}
\newcommand{\ad}[1]{\hat{a}_{\bm{#1}}^\dagger}

\begin{document}

	\title{All coherent field states entangle equally}
	
	\author{Petar Simidzija}
	\affiliation{Department of Applied Mathematics, University of Waterloo, Waterloo, Ontario, N2L 3G1, Canada}
	\affiliation{Institute for Quantum Computing, University of Waterloo, Waterloo, Ontario, N2L 3G1, Canada}
	
	\author{Eduardo Mart\'in-Mart\'inez}
	\affiliation{Department of Applied Mathematics, University of Waterloo, Waterloo, Ontario, N2L 3G1, Canada}
	\affiliation{Institute for Quantum Computing, University of Waterloo, Waterloo, Ontario, N2L 3G1, Canada}
	\affiliation{Perimeter Institute for Theoretical Physics, Waterloo, Ontario N2L 2Y5, Canada}

	\begin{abstract}
	We analyze the interactions of particle detectors with coherent states of a free scalar field. We find that the eigenvalues of the post-interaction density matrices of i) a single detector, ii) two detectors, and iii) the partial transpose of the latter, are all independent of which coherent state the field was in. A consequence of these results is that a detector pair can harvest the same amount of entanglement from any coherent field state as from the vacuum.
	\end{abstract}
	
	\maketitle
	
	\section{Introduction}\label{sec:intro}
	
	The study of entanglement in the context of quantum field theory has garnered a lot of attention over the last several decades. In particular, since Summers and Werner showed that the vacuum state of a free quantum field is entangled~\cite{Summers1985,Summers1987}, there has been a plethora of work seeking to discover operational approaches that make use of this entanglement, such as, for example, in protocols of \textit{quantum energy teleportation}~\cite{Hotta2008,Hotta2009,Frey2014}.
	
	Another protocol, \textit{entanglement harvesting}, was first studied by Valentini~\cite{Valentini1991} and later by Reznik \textit{et al.}~\mbox{\cite{Reznik2003,Reznik2005}}. They showed that a pair of quantum two-level systems (which we call particle detectors) that are initially uncorrelated, can become entangled after locally interacting with the background quantum field. In fact the detectors do not even need to be causally connected for this to occur, a result that operationally proves that spacelike separated regions of a quantum field are entangled.
	
	Since these early studies first investigated entanglement harvesting from the scalar vacuum, it has been found that entanglement harvesting is also possible in timelike separation~\cite{Braun2002,Braun2005,Olson2012} from thermal states~\cite{Braun2005,Brown2013a}, as well as from the electromagnetic vacuum using fully-equipped hydrogen-like atoms~\cite{Pozas2016}. Designing experiments to realize these protocols is, in principle, within reach of current technology. For instance, the experimental feasibility of entanglement harvesting, both timelike and spacelike, has already been   assessed in previous literature~\cite{Olson2011,Sabin2012,Brown2015}.
	
	The study of entanglement harvesting is of interest for both fundamental and practical reasons. For example, on the fundamental side, entanglement harvesting is highly sensitive to spacetime geometry~\cite{Steeg2009} and topology~\cite{Martinez2016a}. On the more applied side (but still in a theoretical context), possible applications of entanglement harvesting have been proposed in the field of metrology, such as in rangefinding~\cite{Salton2015} and precise vibration detection~\cite{Brown2014}. Meanwhile, others have suggested that it may be possible to repeatedly harvest and distill significant amounts of entanglement from the quantum field into atomic-based Bell pairs, which could then be used for quantum information purposes~\cite{Martinez2013a}.
	
To deepen our fundamental understanding of the phenomenon of entanglement harvesting, and to take the step from theoretical proposals towards an experimental realization of an entanglement harvesting protocol, it is necessary to investigate what parameters can be altered in the physical setup so as to optimize the amount of harvestable entanglement. The dependence of entanglement harvesting from the scalar vacuum on relevant parameters of the setup, such as spacetime dimensionality and the properties of the detectors (e.g. their energy gaps), has been investigated in some detail in past literature ~\cite{Pozas2015,pozas2017}. However, excepting a few cases such as thermal states, there have not been any studies of the field state dependence of entanglement harvesting known to the authors.
	
    Thinking along these lines, we ask the following question: if we shine two Unruh-DeWitt particle detectors with coherent light (or more precisely, if we allow them to interact with a coherent field state), then how much entanglement can they harvest from the field? In answering this question we will come to a much more general result. Namely, we will show that following the interaction of the detectors with the field, the one- and two-detector density matrix eigenvalues, as well as the eigenvalues of the partially-transposed two-detector density matrix, are independent of which coherent state the field was in, at least to second order in the detectors' coupling strengths to the field. This means that any property of the detectors that depends only on the spectra of these matrices (e.g., the von Neumann entropy) will be invariant under phase-space displacements of the vacuum. Hence, as a corollary, we will answer our question: irrespective of the detectors' properties or the spacetime dimensionality, the detector pair can harvest the same amount of entanglement from any coherent state as it can from the vacuum.	
	
	This paper is organized as follows. In Sec.~\ref{sec:setup_coherent_states} we introduce coherent states of a free scalar field. Then, in Sec.~\ref{sec:setup_interaction}, we review the concept of particle-detectors and describe their interactions with the field using the Unruh-DeWitt model. In Sec.~\ref{sec:results} we calculate density matrix and partially-transposed density matrix spectra. We present our conclusions in Sec.~\ref{sec:conclusions}. Additional technical details are provided in Appendices~\ref{appendixA} and~\ref{appendixB}, and natural units $\hbar=c=1$ are used throughout.

	\section{Setup}
	\label{sec:setup}
	
	\subsection{Coherent states of the scalar quantum field}\label{sec:setup_coherent_states}

	A scalar quantum field in $(n+1)$-dimensional flat spacetime can be expanded in plane wave modes as
	\begin{equation}\label{eq:field}
		\hat{\phi}(\bm{x},t)=\int\frac{\d[n]{\bm{k}}}{\sqrt{2(2\pi)^n |\bm{k}|}}\left[\ad{k} e^{\ii(|\bm{k}|t-\bm{k}\cdot\bm{x})}+\text{H.c.}\right],
	\end{equation}
	with the creation and annihilation operators, $\ad{k}$ and $\a{k}$, respectively, satisfying the canonical commutation relations
	\begin{equation}\label{eq:CCR}
		 [\a{k},\a{k'}]= [\ad{k},\ad{k'}]=0, \quad
		 [\a{k},\ad{k'}]=\delta^{(n)}(\bm{k}-\bm{k'}).
	\end{equation}
	The ground state of the field is denoted $\ket{0}$ and by definition satisfies
	\begin{equation}\label{eq:ground_state}
		\a{k}\ket{0}=0,
	\end{equation}
	for all wavevectors $\bm{k}\in {\rm I\!R}^n$.
	
	Analogously to coherent states of a simple harmonic oscillator, we define a coherent state of the quantum field to be any state $\ket{\alpha(\bm{k})}$ satisfying
	\begin{equation}\label{eq:coh_1}
		\a{k'}\ket{\alpha(\bm{k})}=\alpha(\bm{k'})\ket{\alpha(\bm{k})}.
	\end{equation}
	The complex valued \textit{coherent amplitude} distribution $\alpha(\bm{k})$ characterizes the coherent state $\ket{\alpha(\bm{k})}$. Notice that the ground state $\ket{0}$, as defined in~\eqref{eq:ground_state}, is a coherent state with vanishing coherent amplitude. We will find it useful to write $\ket{\alpha(\bm{k})}$ as a displacement of the vacuum in phase space:
	\begin{equation}
	    \label{eq:alpha}
		\ket{\alpha(\bm{k})}
		\coloneqq
		\hat{D}_{\alpha(\bm{k})}\ket{0}\coloneqq \exp\!
		\left(\int\!\!\d[n]{\bm{k}}\!\!\left[\alpha(\bm{k})\ad{k}
		\!-\!
		\alpha^*(\bm{k})\a{k}\right]\!\right)
		\!\!\ket{0}.
	\end{equation}
	The proof that this is indeed a coherent state satisfying~\eqref{eq:coh_1} is straightforward. Using the Baker-Campbell-Hausdorff formula~\cite{Truax1988} and the canonical commutation relations~\eqref{eq:CCR} we find that
	\begin{equation}
	\label{eq:commutator}
		[\a{k'},\hat{D}_{\alpha(\bm{k})}]=
		\alpha(\bm{k'})\hat{D}_{\alpha(\bm{k})}.
	\end{equation}
	Together with~\eqref{eq:ground_state} this immediately proves the result.
	
	The unitary operator $\hat{D}_{\alpha(\bm{k})}$ is usually referred to as a displacement operator because it  implements translations in phase space. In other words, the phase space distribution of $\ket{\alpha(\bm{k})}$ is a Gaussian centered away from the origin that saturates the Heisenberg uncertainty relation. Coherent states are particularly relevant in quantum optics since they  model coherent light (e.g. from a laser), for which the photon number statistics follow a Poisson distribution~\cite{scully1999}.
	
	\subsection{Detector-field interaction}
	\label{sec:setup_interaction}
	
	\subsubsection{Single detector}
	\label{sec:one_detector}
	
	We first consider a single, stationary particle detector A at position $\bm{x}_\textsc{a}$ in flat spacetime. This detector couples locally to a scalar quantum field according to the Unruh-DeWitt model~\cite{DeWitt1979}, which succeeds in capturing most of the fundamental features of the light-matter interaction when angular momentum exchange does not play a dominant role in the detector's dynamics~\cite{Martinez2013,Alhambra2014,Pozas2016}. This simple but powerful detector model has been successfully used to analyze information flows in relativistic settings in previous literature~\cite{Jonsson2014,Jonsson2015,Blasco2015,Blasco2015a,Blasco2016,Simidzija2017}. The model considers the detector to be a two-level quantum system (i.e. a qubit), with ground state $\ket{g_\textsc{a}}$, excited state $\ket{e_\textsc{a}}$, and energy gap $\Omega_\textsc{a}$. The detector is, in general, considered to have spatial extent, characterized by the smearing function $F_\textsc{a}(\bm{x})$. The finite size of a detector is particularly relevant when its trajectory is non-inertial~\cite{Schlicht2004,Langlois2004,Louko2006,Martinez2013}. Namely, in order to ensure that the detector keeps its rigid-body form as it accelerates, different points of the detector must experience different accelerations in its center of mass reference frame. However we will not have to deal with these issues, since we will assume the detector to be inertial.
	
	Suppose that the quantum field starts out in the state $\hat{\rho}_{\hat{\phi},0}$, and the detector in the state $\hat{\rho}_{\textsc{a},0}$. Hence the initial state of the detector-field system is
	\begin{equation}
	\label{eq:rho_0_1}
		\hat{\rho}_0
		\coloneqq
		\hat{\rho}_{\textsc{a},0}
		\otimes\hat{\rho}_{\hat{\phi},0}.
	\end{equation}
	We allow this state to evolve according to the the Unruh-DeWitt interaction Hamiltonian, which in the interaction picture takes the form
	\begin{equation}\label{eq:H_I_nu}
		\hat{H}_{\textsc{i},\textsc{a}}(t)=
		\lambda_\textsc{a} \chi_\textsc{a}(t) \hat{m}_\textsc{a}(t)
		\int \d[n]{\bm{x}} F_\textsc{a}(\bm{x}-\bm{x}_\textsc{a}) \hat{\phi}(\bm{x},t).
	\end{equation}
	Here, $\lambda_\textsc{a}$ is the coupling between the detector and the field, assumed to be small relative to the pertinent scales in the problem with the same units (in $(n+1)$-dimensions $\lambda_\textsc{a}$ has units of $[\text{length}]^{(n-3)/2}$). The non-negative real valued function $\chi_\textsc{a}(t)$ characterizes the detector switching, and $t$ is the proper time of the stationary detector. Finally,
	\begin{equation}
	    \label{eq:m_A}
		\hat{m}_\textsc{a}(t)=\hat{\sigma}^+_\textsc{a} e^{\ii\Omega_\textsc{a} t}+
		\hat{\sigma}^-_\textsc{a} e^{-\ii\Omega_\textsc{a} t},
	\end{equation}
	is the monopole moment of the detector. Here, \mbox{$\hat{\sigma}^+_\textsc{a}=\ket{e_\textsc{a}}\bra{g_\textsc{a}}$} and $\hat{\sigma}^-_\textsc{a}=\ket{g_\textsc{a}}\bra{e_\textsc{a}}$ are the $\text{SU}(2)$ raising and lowering operators, respectively.
	
	The detector-field system evolves from the initial state $\hat{\rho}_0$ according to the unitary $\hat{U}$ generated by the interaction Hamiltonian~\eqref{eq:H_I_nu}:
	\begin{equation}
		\hat{U}=\mathcal{T}\exp\left[{-\ii\int_{-\infty}^{\infty}\!\!\!\dif t\, \hat{H}_\textsc{i,a}(t)}\right],
	\end{equation}
	where $\mathcal{T}$ denotes the time ordering operation. The final state of the system is
	\begin{equation}
		\hat{\rho}=
		\hat{U}\hat{\rho}_0 \hat{U}^\dagger.
	\end{equation}
We take a perturbative approach to the problem. The Dyson expansion of $\hat{U}$ for small enough $\lambda_\textsc{a}$ is
	\begin{equation}\label{eq:dyson}
		\hat{U}=\mathds{1}\!
		\underbrace{-\ii\!\int_{-\infty}^\infty\!\!\!\!\!\dif t \hat{H}_\textsc{i,a}(t)} _{\hat{U}^{(1)}}
		\underbrace{-\!\!\int_{-\infty}^{\infty}\!\!\!\!\!\dif t\!
		\int_{-\infty}^t\!\!\!\!\!\dif t' \hat{H}_\textsc{i,a}(t)\hat{H}_\textsc{i,a}(t')}
		_{\hat{U}^{(2)}}
		+\mathcal{O}(\lambda_\textsc{a}^3).
	\end{equation}
	Note that the term $\hat{U}^{(i)}$ contains $i$ factors of $\lambda_\textsc{a}$. Thus the final state of the system can be written as
	\begin{equation}
		\hat{\rho}=
		\hat{\rho}_0+
		\hat{\rho}^{(1)}+
		\hat{\rho}^{(2)}+
		\mathcal{O}(\lambda_\textsc{a}^3),
	\end{equation}
	where
	\begin{align}
		\hat{\rho}^{(1)}&\coloneqq
		\hat{U}^{(1)}\hat{\rho}_0
		+\hat{\rho}_0 \hat{U}^{(1)\dagger},
		\\
		\hat{\rho}^{(2)}&\coloneqq
		\hat{U}^{(2)}\hat{\rho}_0
		+\hat{U}^{(1)} \hat{\rho}_0 
		\hat{U}^{(1)\dagger}
		+\hat{\rho}_0 \hat{U}^{(2)\dagger}.
	\end{align}
	The final state $\hat{\rho}_{\textsc{a}}$ of the detector subsystem is then obtained by tracing out the field. Namely
	\begin{equation}
	    \label{eq:rho_A1}
		\hat{\rho}_\textsc{a}
		\coloneqq
		\Tr_{\hat{\phi}}(\hat{\rho})=
		\hat{\rho}_{\textsc{a},0}+
		\hat{\rho}_\textsc{a}^{(1)}+
		\hat{\rho}_\textsc{a}^{(2)}
		+\mathcal{O}(\lambda_\textsc{a}^3),
	\end{equation}
	where
	\begin{equation}
		\hat{\rho}_\textsc{a}^{(i)}
		\coloneqq
		\Tr_{\hat{\phi}}(\hat{\rho}^{(i)}).
	\end{equation}
    
From~\eqref{eq:dyson} and~\eqref{eq:H_I_nu}, one obtains the following expressions for the leading order contributions to the time evolved density matrix:
	\begin{align}
	\label{eq:rho_nu_1}
		\hat{\rho}_\textsc{a}^{(1)}&=			\ii \lambda_\textsc{a}\int_{-\infty}^{\infty}\!\!\!\!\!\dif t 
		\chi_\textsc{a}(t)
		[\hat{\rho}_{\textsc{a},0},\hat{m}_\textsc{a}(t)]
		V(\bm{x}_\textsc{a},t), \\
	\label{eq:rho_nu_2}
		\hat{\rho}_\textsc{a}^{(2)}&=
		\lambda_\textsc{a}^2
		\Bigg[\int_{-\infty}^{\infty}\!\!\!\!\!\dif t			\int_{-\infty}^{\infty}\!\!\!\!\!\dif t'
		\chi_\textsc{a}(t')\chi_\textsc{a}(t) \notag\\
		&\phantom{{}=
				\!\!\!\!\sum_{\nu,\eta\in\{\textsc{a,b}\}}\!\!\!\!
				\lambda_\nu}
		\times
		\hat{m}_\textsc{a}(t')\hat{\rho}_{\textsc{a},0}
		\hat{m}_\textsc{a}(t) 
		W(\bm{x}_\textsc{a},t,\bm{x}_\textsc{a},t')\notag\\
			&\phantom{{}=\,\,}
			-\int_{-\infty}^{\infty}\!\!\!\!\!\dif t
			\int_{-\infty}^{t}\!\!\!\!\!\dif t'
			\chi_\textsc{a}(t)\chi_\textsc{a}(t') \notag\\
			&\phantom{{}=
				\!\!\!\!\sum_{\nu,\eta\in\{\textsc{a,b}\}}\!\!\!\!
				\lambda_\nu}
			\times\hat{m}_\textsc{a}(t)\hat{m}_\textsc{a}(t')
			\hat{\rho}_{\textsc{a},0}
			W(\bm{x}_\textsc{a},t,\bm{x}_\textsc{a},t')
			\notag\\
			&\phantom{{}=\,\,}
			-\int_{-\infty}^{\infty}\!\!\!\!\!\dif t
			\int_{-\infty}^{t}\!\!\!\!\!\dif t'
			\chi_\textsc{a}(t)\chi_\textsc{a}(t') \notag\\
			&\phantom{{}=
				\!\!\!\!\sum_{\nu,\eta\in\{\textsc{a,b}\}}\!\!\!\!
				\lambda_\nu}
			\times
			\hat{\rho}_{\textsc{a},0}\hat{m}_\textsc{a}(t')\hat{m}_\textsc{a}(t)
			W(\bm{x}_\textsc{a},t',\bm{x}_\textsc{a},t)
		\Bigg],
	\end{align}
	where $V(\bm{x}_\textsc{a},t)$ and $W(\bm{x}_\textsc{a},t,\bm{x}_\textsc{a},t')$ are given by
	\begin{align}
	\label{eq:V}
		V(\bm{x}_\textsc{a},t)
		&\coloneqq
		\int\d[n]{\bm{x}}
		F_\textsc{a}(\bm{x}-\bm{x}_\textsc{a})
			v(\bm{x},t),\\
	\label{eq:W}
		W(\bm{x}_\textsc{a},t,\bm{x}_\textsc{a},t') 
		&\coloneqq
		\int\!\!\d[n]{\bm{x}}\!\!\!\int\!\!\d[n]{\bm{x'}}
			F_\textsc{a}(\bm{x}-\bm{x}_\textsc{a}) F_\textsc{a}(\bm{x'}-\bm{x}_\textsc{a})
			\notag\\
			&\phantom{{}=}\times w(\bm{x},t,\bm{x'},t'),
	\end{align}
	and they denote the pullbacks of the one- and two-point functions, $v(\bm{x},t)$ and $w(\bm{x},t,\bm{x'},t')$, respectively (the latter one also known as the Wightman function), on the detectors' smeared worldlines. These functions are defined as
	\begin{align}
	    \label{eq:v}
		v(\bm{x},t)
		&\coloneqq
		\Tr_{\hat{\phi}}
		[\hat{\phi}(\bm{x},t)\rho_{\hat{\phi},0}], 
		\\
		\label{eq:w}
		w(\bm{x},t,\bm{x'},t')
		&\coloneqq
		\Tr_{\hat{\phi}}
			[\hat{\phi}(\bm{x},t)
			\hat{\phi}(\bm{x'},t')
			\rho_{\hat{\phi},0}],
	\end{align}
	for an arbitrary initial field state $\hat{\rho}_{\hat\phi,0}$.
	
	\subsubsection{Two detectors}
	
Let us now consider the case of \textit{two} two-level particle detectors (labelled \mbox{$\nu\in\{\text{A}, \text{B}\}$}), with ground and excited states $\ket{g_\nu}$ and $\ket{e_\nu}$, energy gaps $\Omega_\nu$, and spatial smearing functions $F_\nu(\bm{x})$. We consider the detectors to be at rest with their centers of mass at positions $\bm{x}_\nu$ in flat spacetime. Let us denote the initial state of the two detectors by $\hat{\rho}_{\textsc{ab},0}$, in which case the initial state of the detectors-field system is 
	\begin{equation}
	    \label{eq:rho_0_2}
		\hat{\rho}_0=\hat{\rho}_{\textsc{ab},0}
		\otimes\hat{\rho}_{\hat{\phi},0}.
	\end{equation}
    Again we assume that the system evolves according to the Unruh-DeWitt interaction Hamiltonian, which in the interaction picture takes the form
	\begin{equation}\label{eq:udw_hamiltonian}
		\hat{H}_\textsc{i,ab} (t)=\!\!\!\sum_{\nu\in\{\text{A,B}\}}
		\!\!\!\lambda_\nu \chi_\nu(t) \hat{\mu}_\nu(t)
		\int \d[n]{\bm{x}} F_\nu(\bm{x}-\bm{x}_\nu) \hat{\phi}(\bm{x},t).
	\end{equation}
	Here $\chi_\nu(t)$ is the switching function for detector $\nu$ and
	\begin{equation}
	\label{eq:monopole}
		\hat{\mu}_\textsc{a}(t)
		\coloneqq
		\hat{m}_\textsc{a}(t)\otimes\mathds{1}_\textsc{b},\quad 
		\hat{\mu}_\textsc{b}(t)
		\coloneqq
		\mathds{1}_\textsc{a}\otimes\hat{m}_\textsc{b}(t)
	\end{equation}
	are operators acting on the two-detector Hilbert space $\mathcal{H}_\textsc{a}\otimes\mathcal{H}_\textsc{b}$, with $\hat{m}_\nu$ given by
	\begin{equation}
	    \label{eq:m_nu}
		\hat{m}_\nu(t)=\hat{\sigma}^+_\nu e^{\ii\Omega_\nu t}+
		\hat{\sigma}^-_\nu e^{-\ii\Omega_\nu t}.
	\end{equation}

	Following the same approach as in the single-detector case, we find that the state of the two detectors after their interaction with the field is
	\begin{equation}
	    \label{eq:rho_ab}
	    \hat{\rho}_\textsc{ab}=
	    \hat{\rho}_{\textsc{ab},0}+
	    \hat{\rho}_\textsc{ab}^{(1)}+ \hat{\rho}_\textsc{ab}^{(2)} + \mathcal{O}(\lambda_\nu^3).
	\end{equation}
	Here, we define $\mathcal{O}(\lambda_\nu^3)$ to mean $\mathcal{O}(\lambda_\textsc{a}^i\lambda_\textsc{b}^j)$ with $(i,j)\in\{(0,2),(1,1),(2,0)\}$. The terms $\hat{\rho}_\textsc{ab}^{(1)}$ and $\hat{\rho}_\textsc{ab}^{(2)}$ are given by
	\begin{align}
	\label{eq:rho_ab_1}
	\hat{\rho}_{\textsc{ab}}^{(1)}&=	\ii\!\!\!\!\sum_{\nu\in\{\text{A,B}\}}\!\!\!\!\lambda_\nu
		\int_{-\infty}^{\infty}\!\!\!\!\!\dif t\chi_\nu(t)
		[\hat{\rho}_{\textsc{ab},0},\hat{\mu}_\nu(t)]V(\bm{x}_\nu,t), 
	\\
	\label{eq:rho_ab_2}
		\hat{\rho}_{\textsc{ab}}^{(2)}&=
		\!\!\!\!\sum_{\nu,\eta\in\{\text{A,B}\}}\!\!\!\!
		\lambda_\nu\lambda_\eta
		\Bigg[\int_{-\infty}^{\infty}\!\!\!\!\!\dif t
			\int_{-\infty}^{\infty}\!\!\!\!\!\dif t'
			\chi_\nu(t')\chi_\eta(t) \notag\\
			&\phantom{{}=
				\!\!\!\!\sum_{\nu,\eta\in\{\textsc{a,b}\}}\!\!\!\!
				\lambda_\nu}
			\times\hat{\mu}_\nu(t')
			\hat{\rho}_{\textsc{ab},0}\hat{\mu}_\eta(t)
			W(\bm{x}_\eta,t,\bm{x}_\nu,t')\notag\\
			&\phantom{{}=\,\,}
			-\int_{-\infty}^{\infty}\!\!\!\!\!\dif t
			\int_{-\infty}^{t}\!\!\!\!\!\dif t'
			\chi_\nu(t)\chi_\eta(t') \notag\\
			&\phantom{{}=
				\!\!\!\!\sum_{\nu,\eta\in\{\textsc{a,b}\}}\!\!\!\!
				\lambda_\nu}
			\times\hat{\mu}_\nu(t)\hat{\mu}_\eta(t')
			\hat{\rho}_{\textsc{ab},0}
			W(\bm{x}_\nu,t,\bm{x}_\eta,t')\notag\\
			&\phantom{{}=\,\,}
			-\int_{-\infty}^{\infty}\!\!\!\!\!\dif t
			\int_{-\infty}^{t}\!\!\!\!\!\dif t'
			\chi_\nu(t)\chi_\eta(t') \notag\\
			&\phantom{{}=
				\!\!\!\!\sum_{\nu,\eta\in\{\textsc{a,b}\}}\!\!\!\!
				\lambda_\nu}
			\times
			\hat{\rho}_{\textsc{ab},0}\hat{\mu}_\eta(t')\hat{\mu}_\nu(t)
			W(\bm{x}_\eta,t',\bm{x}_\nu,t)
		\Bigg],
	\end{align}
	where $V(\bm{x}_\nu,t)$ and $W(\bm{x}_\eta,t,\bm{x}_\nu,t')$ are defined as	
	\begin{align}
	\label{eq:V_2}
		V(\bm{x}_\nu,t)
		&\coloneqq
		\int\d[n]{\bm{x}}
		F_\nu(\bm{x}-\bm{x}_nu)
			v(\bm{x},t),\\
	\label{eq:W_2}
		W(\bm{x}_\eta,t,\bm{x}_\nu,t') &\coloneqq
		\int\!\!\d[n]{\bm{x}}\!\!\!\int\!\!\d[n]{\bm{x'}}
			F_\eta(\bm{x}-\bm{x}_\eta) F_\nu(\bm{x'}-\bm{x}_\nu)
			\notag\\
			&\phantom{{}=}\times w(\bm{x},t,\bm{x'},t'),
	\end{align}
	and the one- and two-point functions $v(\bm{x},t)$ and $w(\bm{x},t,\bm{x'},t')$ are given in~\eqref{eq:v} and~\eqref{eq:w}, respectively.

	\section{Results}
	\label{sec:results}
	
	Thus far the only restriction we have placed on the initial state of the detector(s)-field system is that the field and detectors are uncorrelated, i.e., the joint state can be written in the  form~\eqref{eq:rho_0_1} (one detector) or~\eqref{eq:rho_0_2} (two detectors). We now set the initial state of the field to be a coherent state $\ket{\alpha(\bm{k})}$, as defined in~\eqref{eq:alpha}, so that
	\begin{equation}
	    \label{eq:rho_phi}
	    \hat{\rho}_{\hat{\phi},0}=
	    \ket{\alpha(\bm{k})}\bra{\alpha(\bm{k})}.
	\end{equation}
	We can now  compute the one- and two-point  functions $v(\bm{x},t)$ and $w(\bm{x},t,\bm{x'},t')$, defined in~\eqref{eq:v} and~\eqref{eq:w}, respectively. For $v(\bm{x},t)$ we obtain
	\begin{align}
	    \label{eq:v_coh}
	    v&(\bm{x},t) \\
	    &=\bra{\alpha(\bm k)}
	    \hat{\phi}(\bm{x},t)
	    \ket{\alpha(\bm k)} \notag\\
	    &= \bra{0}\hat{D}_{\alpha(\bm k)}^\dagger
	    \hat{\phi}(\bm{x},t)
	    \hat{D}_{\alpha(\bm k)}\ket{0}
	    \notag\\
	    &=
	    \bra{0}\hat{D}_{\alpha(\bm k)}^\dagger\!\!\!
	    \int\!\!\!\frac{\d[n]{\bm{k}_1}}{\!\!\sqrt{2(2\pi)^n |\bm{k}_1|}}\!\!\left[\ad{k_1}\! e^{\ii(|\bm{k}_1|t-\bm{k}_1\cdot\bm{x})}\!\!+\!\text{H.c.}\!\right]\!\!
	    \hat{D}_{\alpha(\bm k)}\ket{0}
	    \notag\\
	    &=	    
	    \int\!\!
	    \frac{\d[n]{\bm{k}}}{\sqrt{2(2\pi)^n|\bm{k}|}}
	    (\alpha^*(\bm{k})e^{\ii(|\bm{k}|t-\bm{k}\cdot\bm{x})}
	    +\text{H.c.}),\notag
	\end{align}
	where in the last line we used the commutator~\eqref{eq:commutator}. Similarly, the Wightman function $w(\bm{x},t,\bm{x'},t')$ is
	\begin{align}
	    w&(\bm{x},t,\bm{x}',t')\notag\\
	    &= \bra{0}\hat{D}_{\alpha(\bm k)}^\dagger
	    \hat{\phi}(\bm{x},t)
	    \hat{\phi}(\bm{x'},t')
	    \hat{D}_{\alpha(\bm k)}\ket{0}
	    \notag\\
	    &= \bra{0}\hat{D}_{\alpha(\bm k)}^\dagger 
	    \int\!\!\frac{\d[n]{\bm{k}_1}}{\sqrt{2(2\pi)^n |\bm{k}_1|}}\left[\ad{k_1} e^{\ii(|\bm{k}_1|t-\bm{k}_1\cdot\bm{x})}+\text{H.c.}\right] \notag\\
	    &\phantom{= }\times\!
	    \int\!\!\frac{\d[n]{\bm{k}_2}}{\sqrt{2(2\pi)^n |\bm{k}_2|}}\left[\ad{k_2} e^{\ii(|\bm{k}_2|t'-\bm{k}_2\cdot\bm{x'})}+\text{H.c.}\right]
	    \!\hat{D}_{\alpha(\bm k)}\ket{0}.
    \end{align}
    Using~\eqref{eq:commutator} we can simplify the above expression to obtain
    \begin{align}
	    w(\bm{x},t,\bm{x}',t')&=
	    J(\bm{x},t)J(\bm{x'},t')+
	    J(\bm{x},t)J(\bm{x'},t')^*
	    \notag\\
	    &\phantom{=}+
	    J(\bm{x},t)^*J(\bm{x'},t')+
	    J(\bm{x},t)^*J(\bm{x'},t')^*
	    \notag\\
	    &\phantom{=}+
	    w_\text{vac}(\bm{x},t,\bm{x}',t').
    \end{align}
	Here, $J(\bm{x},t)$ and $w_\text{vac}(\bm{x},t,\bm{x}',t')$ are given by
	\begin{align}
	    \label{eq:J}
	    J(\bm{x},t)
	    &\coloneqq\!
	    \int\!\frac{\d[n]{\bm{k}}}{\sqrt{2(2\pi)^n |\bm{k}|}}
	    \alpha(\bm k) e^{-\ii(|\bm{k}|t-\bm{k}\cdot\bm{x})},\\
	    \label{eq:w_vac}
	    w_\text{vac}(\bm{x},t,\bm{x}'\!,t')\!
	    &\coloneqq\! \int\!\frac{\d[n]{\bm{k}}}{2(2\pi)^n|\bm{k}|}
	    e^{-\ii(|\bm{k}|t-\bm{k}\cdot\bm{x})}
	    e^{\ii(|\bm{k}|t'-\bm{k}\cdot\bm{x'})}.
	\end{align}
	Note that, since $w_\text{vac}(\bm{x},t,\bm{x}',t')$ is independent of the coherent amplitude of the field, $\alpha(\bm{k})$, this term is present even when $\alpha(\bm k)$ is identically zero. In fact, $w_\text{vac}(\bm{x},t,\bm{x}',t')$ corresponds exactly to the Wightman function of the vacuum state of the field---we call it a \textit{vacuum term}. On the other hand, $J(\bm{x},t)$ \textit{does} depend on $\alpha(\bm{k})$, and vanishes when \mbox{$\alpha(\bm k)\coloneqq0$}. It is easy to verify that $v(\bm{x},t)=2\Realpart[J(\bm{x},t)]$, and hence we can write $w(\bm{x},t,\bm{x'},t')$ as:
	\begin{equation}
	    \label{eq:w_coh}
	    w(\bm{x},t,\bm{x}'\!,t')=
	    v(\bm{x},t)v(\bm{x'},t')
	    +w_\text{vac}(\bm{x},t,\bm{x}',t').
	\end{equation}
	Notice that the two-point function is a sum of the vacuum Wightman function and a product of two $\alpha(\bm k)$-dependent one-point functions.

	We can now compute $V(\bm{x}_\nu,t)$ and $W(\bm{x}_\eta,t,\bm{x}_\nu,t')$ from~\eqref{eq:V_2} and~\eqref{eq:W_2}, respectively, obtaining
	\begin{align}
	    \label{eq:V_coh}
	    V(\bm{x}_\nu,t)&=
		\int\d[n]{\bm{x}} F_\nu(\bm{x}-\bm{x}_\nu)
			v(\bm{x},t),\\
	    \label{eq:W_coh}
		W(\bm{x}_\nu,t,\bm{x}_\eta,t') &=
		V(\bm{x}_\nu,t) V(\bm{x}_\eta,t')+
		W_\text{vac}(\bm{x}_\nu,t,\bm{x}_\eta,t').
	\end{align}
	The vacuum term $W_\text{vac}(\bm{x}_\nu,t,\bm{x}_\eta,t')$ is the pullback of $w_\text{vac}(\bm{x},t,\bm{x}',t')$ on the detectors' smeared worldlines:
	\begin{align}
	    W_\text{vac}(\bm{x}_\nu,t,\bm{x}_\eta,t')
	    &\coloneqq
		\int\!\!\d[n]{\bm{x}}\!\!\!\int\!\!\d[n]{\bm{x'}}
			F_\nu(\bm{x}-\bm{x}_\nu) F_\eta(\bm{x'}-\bm{x}_\eta)
			\notag\\
			&\phantom{{}=}\times w_\text{vac}(\bm{x},t,\bm{x'},t').
	\end{align}
	To proceed with the calculations of the one- and two-detector final state density matrices, we must now set the initial conditions for the detectors.
	
	\subsection{Single detector}
	
	Let us suppose that a single detector A starts out in its ground state:
	\begin{equation}
	    \label{eq:rho_A_0}
	    \hat{\rho}_{\textsc{a},0}=
	    \ket{g_\textsc{a}}\bra{g_\textsc{a}}.
	\end{equation}
	Then, from~\eqref{eq:rho_nu_1} and~\eqref{eq:rho_nu_2}, the state of the detector following its interaction with the field is (in the $\{\ket{g_\textsc{a}},\ket{e_\textsc{a}}\}$ basis)
	\begin{equation}
	\label{eq:rho_A}
		\hat{\rho}_\textsc{a}=
		\begin{pmatrix}
		1-\mathcal{L}_{\textsc{aa}}-
		\bar{\mathcal{L}}_{\textsc{aa}} & \bar{L}_\textsc{a}^*\\
		\bar{L}_\textsc{a} & \mathcal{L}_{\textsc{aa}}+
		\bar{\mathcal{L}}_{\textsc{aa}}
		\end{pmatrix}
		+\mathcal{O}(\lambda_\textsc{a}^3),
	\end{equation}
	where we have denoted with an overbar the terms that explicitly depend on the coherent amplitude of the field (and vanish for $\alpha(\bm k)\coloneqq0$). The terms without an overbar are present even when the field starts in the vacuum state. The vacuum terms are \cite{Pozas2015}:
	\begin{equation}
	    \label{eq:L_AA}
	    \mathcal{L}_{\textsc{aa}}
	    \coloneqq
		\int\d[n]{\bm{k}}L_\textsc{a}(\bm{k})L_\textsc{a}(\bm{k})^*,
	\end{equation}
	with $L_\textsc{a}(\bm{k})$ defined by
	\begin{equation}
	    \label{eq:L_A}
	    L_\textsc{a}(\bm{k})
	    \coloneqq
		\lambda_\textsc{a}
		\frac{e^{-\ii\bm{k}\cdot\bm{x}_\textsc{a}}
		\tilde{F}_\textsc{a}(\bm{k})}{\sqrt{2|\bm{k}|}}
		\int_{-\infty}^{\infty}\!\!\!\!\!\dif t \,\chi_\textsc{a}(t)
		e^{\ii(|\bm{k}|+\Omega_\textsc{a})t}.
	\end{equation}
	Here, $\tilde{F}_\textsc{a}(\bm{k})$ is the Fourier transform of the detector's smearing function $F_\textsc{a}(\bm{x})$,
	\begin{equation}
	    \label{eq:FT}
		\tilde{F}_\textsc{a}(\bm{k})
		\coloneqq
		\frac{1}{\sqrt{(2\pi)^n}}\int\d[n]{\bm{x}}
		F_\textsc{a}(\bm{x})e^{\ii\bm{k}\cdot\bm{x}}.
	\end{equation}
	The $\alpha(\bm k)$-dependent terms in~\eqref{eq:rho_A}, $\bar{\mathcal{L}}_{\textsc{aa}}$ and $\bar{L}_\textsc{a}$, are
	\begin{align}
	    \label{eq:L_AA_bar}
		\bar{\mathcal{L}}_{\textsc{aa}}
		&\coloneqq
		\bar{L}_\textsc{a}\bar{L}_\textsc{a}^*,
		\\
		\label{eq:L_A_bar}
		\bar{L}_\textsc{a}
		&\coloneqq
		-\ii\lambda_\textsc{a}
		\int_{-\infty}^{\infty}\!\!\!\!\dif t			\chi_\textsc{a}(t)e^{\ii\Omega_\textsc{a} t}V(\bm{x}_\textsc{a},t),
	\end{align}
	with $V(\bm{x}_\textsc{a},t)$ given in~\eqref{eq:V_coh} for a coherent state.
		
	Let us investigate the density matrix $\hat{\rho}_\textsc{a}$ (eq.~\eqref{eq:rho_A}) in more detail. Comparing it to the initial-time density matrix $\hat{\rho}_{\textsc{a},0}$ (eq.~\eqref{eq:rho_A_0}), we find that the probability of measuring the detector in the excited state increases from the initial value of $\bra{e_\textsc{a}} \hat{\rho}_{\textsc{a},0} \ket{e_\textsc{a}}=0$ to $\bra{e_\textsc{a}} \hat{\rho}_\textsc{a} \ket{e_\textsc{a}} = \mathcal{L}_{\textsc{aa}}+		\bar{\mathcal{L}}_{\textsc{aa}}$. These field-induced excitations are partly due to the vacuum term $\mathcal{L}_{\textsc{aa}}$, as well as the $\alpha(\bm k)$-dependent term $\bar{\mathcal{L}}_{\textsc{aa}}$, both of which are real and positive. The fact that a non-vanishing coherent amplitude increases the detector's excitation probability is not surprising if we recall that coherent states (of the electromagnetic field) describe coherent light \cite{scully1999}. Indeed, the odds of finding a detector in its excited state increase if we shine it with a laser. 
	
	Next let us calculate the eigenvalues $E_{\textsc{a},i}$ of the time evolved density matrix $\hat{\rho}_\textsc{a}$ of detector A~\eqref{eq:rho_A}. We find that its eigenvalues are
	\begin{align}
	    \label{eq:1_detector_evalues}
	    E_{\textsc{a},1}&=1- \mathcal{L}_{\textsc{aa}}+
	    \mathcal{O}(\lambda_\textsc{a}^3), \\
	    E_{\textsc{a},2}&=\mathcal{L}_{\textsc{aa}}+
	    \mathcal{O}(\lambda_\textsc{a}^3).
	\end{align}
Very remarkably, and unlike the excitation probability, the eigenvalues are independent of the coherent amplitude. Hence we can make the following statement:

\vspace{2mm}
\textbf{Theorem 1:} \textit{Consider an UDW particle detector with arbitrary spatial smearing $F_\textsc{a}(\bm x)$, arbitrary switching function $\chi_\textsc{a}(t)$, and field coupling strength $\lambda_\textsc{a}$. Then, to $\mathcal{O}(\lambda_\textsc{a}^2)$, the eigenvalues of the time evolved density matrix of the detector are the same whether the detector interacts with an arbitrary coherent state of a scalar field or with the vacuum.}

\textit{Therefore, any property of the time evolved state of the  detector that depends solely on the spectrum of its density matrix is independent of whether the detector interacts with an arbitrary coherent state or with the scalar field vacuum.
}
\vspace{2mm}

For example, the von Neumann entropy \mbox{$S\coloneqq-\sum_i E_{\textsc{a},i}\ln(E_{\textsc{a},i})$} is independent of field coherence. Thus, while a detector interacting with a field state of non-vanishing coherent amplitude will experience a higher excitation probability than a second detector interacting with the vacuum, the states of the two detectors will be equally mixed following the interactions.
	
	Next we will show how the eigenvalues of the two-detector final state density matrix $\hat{\rho}_\textsc{ab}$ are also independent of the coherent amplitude---and therefore so is the amount of entanglement that the detectors harvest from the field.
	
	\subsection{Two detectors}
	
	Suppose the pair of detectors A and B is initially in the separable free ground state
	\begin{equation}
	    \label{eq:rho_ab_0}
	    \hat{\rho}_{\textsc{ab},0}=
	    \ket{g_\textsc{a}}\bra{g_\textsc{a}} \otimes
		\ket{g_\textsc{b}}\bra{g_\textsc{b}}.
	\end{equation}
	Following their interaction with the quantum field, their bipartite state $\hat{\rho}_\textsc{ab}$ is obtained from expressions~\eqref{eq:rho_ab_1} and~\eqref{eq:rho_ab_2}. The matrix representation of the time-evolved state of the detectors is
	\begin{widetext}
	\begin{equation}
	    \label{eq:rho_ab_coh}
		\hat{\rho}_\textsc{ab}=
		\begin{pmatrix}
		1-\mathcal{L}_\textsc{aa}-\mathcal{L}_\textsc{bb} -\bar{\mathcal{L}}_\textsc{aa}-\bar{\mathcal{L}}_\textsc{bb} & \bar{L}_\textsc{b}^* & \bar{L}_\textsc{a}^* & \mathcal{M}^* +\bar{\mathcal{M}}^* \\
		\bar{L}_\textsc{b} & \mathcal{L}_\textsc{bb} +\bar{\mathcal{L}}_\textsc{bb} & \mathcal{L}_\textsc{ab}^* +\bar{\mathcal{L}}_\textsc{ab}^* & 0 \\
		\bar{L}_\textsc{a} & \mathcal{L}_\textsc{ab} +\bar{\mathcal{L}}_\textsc{ab} & \mathcal{L}_\textsc{aa} +\bar{\mathcal{L}}_\textsc{aa} & 0 \\
		\mathcal{M} +\bar{\mathcal{M}} & 0 & 0 & 0
		\end{pmatrix}
		+\mathcal{O}(\lambda_\nu^3),
	\end{equation}
	\end{widetext}
	in the basis
	\begin{equation}
	\label{eq:basis}
	    \{
	\ket{g_\textsc{a}}\otimes\ket{g_\textsc{b}},
	\ket{g_\textsc{a}}\otimes\ket{e_\textsc{b}},
	\ket{e_\textsc{a}}\otimes\ket{g_\textsc{b}},
	\ket{e_\textsc{a}}\otimes\ket{e_\textsc{b}}\}.
	\end{equation}
	As in the one-detector case, overbars indicate terms that depend on the coherent amplitude of the field, and vanish if the field is initialized to the vacuum state. On the other hand, terms without an overbar (the vacuum terms) are present even if the initial field state is the vacuum, as was shown in~\cite{Pozas2015}. The vacuum terms are \cite{Pozas2015}:
	\begin{align}
	    \label{eq:L_mu_nu}
	    \mathcal{L}_{\mu\nu}
	    &\coloneqq
		\int\d[n]{\bm{k}}L_\mu(\bm{k})L_\nu(\bm{k})^*,\\
    	\label{eq:M}
		\mathcal{M}
		&\coloneqq
		\int\d[n]{\bm{k}}M(\bm{k}),
	\end{align}
	where $L_\mu(\bm{k})$ and $M(\bm{k})$ are given by
	\begin{align}
	    L_\nu(\bm{k})
	    &\coloneqq
			\lambda_\nu\frac{e^{-\ii\bm{k}\cdot\bm{x}_\nu}
			\tilde{F}_\nu(\bm{k})}{\sqrt{2|\bm{k}|}}
			\int_{-\infty}^{\infty}\!\!\!\!\dif t \chi_\nu(t)
			e^{\ii(|\bm{k}|+\Omega_\nu)t},
		\\
		M(\bm{k})
		&\coloneqq
		-\frac{\lambda_\textsc{a}\lambda_\textsc{b}}{2|\bm{k}|}
		\int_{-\infty}^{\infty}
		\!\!\!\dif t
		\int_{-\infty}^{t}
		\!\!\!\dif t'
		e^{-\ii|\bm{k}|(t-t')}\notag\\
		&\phantom{{}=}
		\Big[
		\chi_\textsc{a}(t)
		\chi_\textsc{b}(t')
		e^{\ii(\Omega_\textsc{a}t
		+\Omega_\textsc{b}t')}
		e^{\ii\bm{k}\cdot(\bm{x}_\textsc{a}-\bm{x}_\textsc{b})}
		\tilde{F}_\textsc{a}(\bm k)
		\tilde{F}_\textsc{b}(\bm k)^*
		\notag\\
		&\phantom{=}+\!
		\chi_\textsc{b}(t)
		\chi_\textsc{a}(t')
		e^{\ii(\Omega_\textsc{b}t
		+\Omega_\textsc{a}t')}
		e^{\ii\bm{k}\cdot(\bm{x}_\textsc{b}-\bm{x}_\textsc{a})}
		\tilde{F}_\textsc{b}(\bm k)
		\tilde{F}_\textsc{a}(\bm k)^*
		\Big].
	\end{align}
	Recall that $\tilde{F}_\nu(\bm{k})$ is the Fourier transform of the smearing function $F_\nu(\bm{x})$, as defined in~\eqref{eq:FT}.
	The $\alpha(\bm k)$-dependent terms in the density matrix $\hat{\rho}_{\textsc{ab}}$ are (see Appendix~\ref{appendixA})
	\begin{align}
	    \label{eq:L_mu_nu_bar}
		\bar{\mathcal{L}}_{\mu\nu}
		&\coloneqq
		\bar{L}_\mu\bar{L}_\nu^*,
		\\
		\label{eq:M_bar}
		\bar{\mathcal{M}}
		&\coloneqq
		\bar{L}_\textsc{a}
		\bar{L}_\textsc{b},
		\\
		\label{eq:L_nu_bar}
		\bar{L}_\nu
		&\coloneqq
		-\ii\lambda_\nu
		\int_{-\infty}^{\infty}\!\!\!\!\dif t			\chi_\nu(t)e^{\ii\Omega_\nu t}V(\bm{x}_\nu,t),
	\end{align}
	with $V(\bm{x}_\nu,t)$ given in~\eqref{eq:V_coh}.
	
	Let us analyze in more detail the time-evolved state of the two-detector system, $\hat{\rho}_{\textsc{ab}}$, given in~\eqref{eq:rho_ab_coh}. In particular, we are interested in whether the state of the (initially unentangled) detectors is entangled following their interactions with a coherent field state, and if so, how does the amount of entanglement compare to when the detectors interact with just the vacuum. Physically, we expect that interacting with a non-vacuum coherent state (e.g. shining the detectors with a laser) will increase local noise and therefore decrease the amount of entanglement between the detector pair. 
	
Mathematical intuition may point to the same direction. The \textit{negativity} of $\hat{\rho}_{\textsc{ab}}$---defined as the negative sum of the negative eigenvalues of the partial transpose of $\hat{\rho}_{\textsc{ab}}$ ~\cite{Vidal2002}---is an entanglement monotone that vanishes only for separable states~\cite{Horodecki1996,Peres1996}, and so it is often used to quantify entanglement between the two detectors in harvesting scenarios. The authors of~\cite{Pozas2015} showed that when the field starts in the vacuum state and the detectors in their ground states, the negativity of the final two-detector state is a direct competition between local terms that decrease the amount of entanglement and non-local terms that increase it. Therefore we might naively expect the $\alpha(\bm k)$-dependent $\bar{L}_\nu$ contributions to $\hat{\rho}_\textsc{ab}$, which are of $\mathcal{O}(\lambda_\nu)$ and hence inherently local, to contribute predominantly to noise terms and thus \textit{decrease} the negativity. At the very least, it would certainly be reasonable to expect that the substantial $\alpha(\bm k)$-dependent contributions to $\hat{\rho}_\textsc{ab}$ will alter the negativity, for better or for worse.
		
 Remarkably, as we will now show, these seemingly reasonable  expectations are not correct: to $\mathcal{O}(\lambda_\nu^2)$ the amount of entanglement that can be harvested from the field is \textit{independent} of its coherent amplitude. In fact we will prove a much more general result:
	
\vspace{2mm}

\textbf{Theorem 2:} \textit{
Consider two UDW particle detectors ($\mathrm{A}$ and $\mathrm{B}$)  with arbitrary spatial smearings $F_\nu(\bm x)$, arbitrary switching functions $\chi_\nu(t)$, and field coupling strengths $\lambda_\nu$, where $\nu\in\{\mathrm{A},\mathrm{B} \}$. Then, to $\mathcal{O}(\lambda_\nu^2)$, the eigenvalues of the time evolved density matrix of the two detectors are the same whether the detectors interact with an arbitrary coherent state of a scalar field or with the vacuum. Additionally,  the eigenvalues of the partially-transposed density matrix of the two detectors are also the same whether the detectors interact with an arbitrary coherent state or with the vacuum.}

\textit{Therefore, any property of the time evolved state of the  detectors that depends solely on the spectrum of their density matrix, or of its partial transpose, is independent of whether the detectors interact with an arbitrary coherent state or with the scalar field vacuum.}
\vspace{2mm}
	
\textit{Proof:} We will show that, to $\mathcal{O}(\lambda_\nu^2)$, the eigenvalues $E_{\textsc{ab},i}^{{\text{\textbf{t}}}_\textsc{b}}$ of $\hat{\rho}_\textsc{ab}^{{\text{\textbf{t}}}_\textsc{b}}$ (the partial transpose of $\hat{\rho}_\textsc{ab}$ with respect to detector B) are independent of the coherent amplitude $\alpha(\bm k)$. The proof for the eigenvalues of $\hat{\rho}_\textsc{ab}$ is similar (see Appendix~\ref{appendixB}). The most obvious way to prove this result is to calculate the eigenvalues and expand them to $\mathcal{O}(\lambda_\nu^2)$. This is straightforward to do, but the expressions obtained this way are very cumbersome. Instead we will take a more indirect approach.
	
From \eqref{eq:rho_ab_coh}, in the basis~\eqref{eq:basis} the partially transposed density matrix $\hat{\rho}_\textsc{ab}^{{\text{\textbf{t}}}_\textsc{b}}$ is given by
\begin{widetext}
	\begin{equation}
	    \label{eq:rho_ab_pt}
		\hat{\rho}_\textsc{ab}^{{\text{\textbf{t}}}_\textsc{b}}
		=
		\begin{pmatrix}
		1-\mathcal{L}_\textsc{aa}-\mathcal{L}_\textsc{bb} -\bar{\mathcal{L}}_\textsc{aa}-\bar{\mathcal{L}}_\textsc{bb} & \bar{L}_\textsc{b} & \bar{L}_\textsc{a}^* & \mathcal{L}_\textsc{ab}^* +\bar{\mathcal{L}}_\textsc{ab}^* \\
		\bar{L}_\textsc{b}^* & \mathcal{L}_\textsc{bb} +\bar{\mathcal{L}}_\textsc{bb} & \mathcal{M}^* +\bar{\mathcal{M}}^* & 0 \\
		\bar{L}_\textsc{a} & \mathcal{M} +\bar{\mathcal{M}} & \mathcal{L}_\textsc{aa} +\bar{\mathcal{L}}_\textsc{aa} & 0 \\
		\mathcal{L}_\textsc{ab} +\bar{\mathcal{L}}_\textsc{ab} & 0 & 0 & 0
		\end{pmatrix}
		+\mathcal{O}(\lambda_\nu^3),
	\end{equation}
	\end{widetext}
Being careful with the consistency of our perturbative expansion, the eigenvalues of $\hat{\rho}_\textsc{ab}^{{\text{\textbf{t}}}_\textsc{b}}$ are the roots of the characteristic polynomial
	\begin{align}
		p(x)&=x^4+[-1 +\mathcal{O}(\lambda_\nu^3)]x^3
		+[C_2
		+\mathcal{O}(\lambda_\nu^3)]x^2
		\notag\\
		\label{eq:char_poly}
		&\phantom{=}
		+[C_4
		+\mathcal{O}(\lambda_\nu^5)]x
		+\mathcal{O}(\lambda_\nu^7),
	\end{align}
	with $C_2\propto \lambda_\nu^2$ and $C_4\propto\lambda_\nu^4$ defined as
	\begin{align}
	    \label{eq:C_2}
        C_2&\coloneqq\mathcal{L}_\textsc{aa}+\mathcal{L}_\textsc{bb},
        \\
		C_4&\coloneqq|\mathcal{M}|^2-
			\mathcal{L}_\textsc{aa}\mathcal{L}_\textsc{bb}.
	\end{align}
	Notice that the leading order coefficients of each power of $x$ in $p(x)$ are independent of the coherent amplitude $\alpha(\bm k)$ (the $\bar{\mathcal{L}}_{\mu\nu}= \bar{L}_\mu\bar{L}_\nu^*$, $\bar{\mathcal{M}}=\bar{L}_\textsc{a} \bar{L}_\textsc{b}$, and $\bar{L}_\nu$ terms cancel exactly in the calculation of the determinant leading to $p(x)$). The most general form for a root of $p(x)$ is given by
	\begin{equation}
	    X=X_0+X_1+X_2+
	    \mathcal{O}(\lambda_\nu^3),
	\end{equation}
    where the $X_i$ term is of order $\mathcal{O}[(\lambda_\nu)^i]$. We now substitute $X$ into $p(x)$ and systematically set terms proportional to $(\lambda_\nu)^i$ equal to zero order by order, starting from $i=0$ and moving up to higher orders until we determine each $X_i$. 
    
    First, equating the terms of $p(X)$ proportional to $\lambda_\nu^0$ to zero, we obtain $0=X_0^3(X_0-1)$. Hence $X_0=0$ or $X_0=1$.
    
    \textbf{Case 1}: $X_0=1$. \\
    Equating the $\lambda_\nu^1$ terms in $p(X)$ to zero, we obtain $X_1=0$. The $\lambda_\nu^2$ terms give $X_2=-C_2$. Therefore, using the definition~\eqref{eq:C_2} of $C_2$, the first root $E_{\textsc{ab},1}^{{\text{\textbf{t}}}_\textsc{b}}$ of the characteristic polynomial $p(x)$ is
    \begin{equation}
        E_{\textsc{ab},1}^{{\text{\textbf{t}}}_\textsc{b}}=1-\mathcal{L}_\textsc{aa}- \mathcal{L}_\textsc{bb}+ \mathcal{O}(\lambda_\nu^3).
    \end{equation}
	
	\textbf{Case 2}: $X_0=0$. \\
	In this case there are no terms in $p(X)$ proportional to $\lambda_\nu^1$ or $\lambda_\nu^2$. Equating the terms proportional to $\lambda_\nu^3$ to zero gives $X_1=0$. There are no terms proportional to $\lambda_\nu^4$ or $\lambda_\nu^5$. Equating the terms proportional to $\lambda_\nu^6$ to zero gives $0=-X_2(X_2^2-C_2X_2-C_4)$. Hence $X_2=0$ or $X_2=(C_2\pm\sqrt{C_2^2+4C_4})/2$. Therefore the remaining roots of $p(x)$ are
	\begin{align}
	    E_{\textsc{ab},2}^{{\text{\textbf{t}}}_\textsc{b}}&=0+\mathcal{O}(\lambda_\nu^3),
	    \\
	    E_{\textsc{ab},3}^{{\text{\textbf{t}}}_\textsc{b}}&=\frac{1}{2}\left(
		\mathcal{L}_\textsc{aa}+\mathcal{L}_\textsc{bb}+
		\sqrt{(\mathcal{L}_\textsc{aa}-\mathcal{L}_\textsc{bb})^2+
		4|\mathcal{M}|^2}\right)
		\notag\\
		&\phantom{=}+\mathcal{O}(\lambda_\nu^3),
		\\
		E_{\textsc{ab},4}^{{\text{\textbf{t}}}_\textsc{b}}&=\frac{1}{2}\left(
		\mathcal{L}_\textsc{aa}+\mathcal{L}_\textsc{bb}-
		\sqrt{(\mathcal{L}_\textsc{aa}-\mathcal{L}_\textsc{bb})^2+
		4|\mathcal{M}|^2}\right)
		\notag \\
		&\phantom{=}+\mathcal{O}(\lambda_\nu^3).
	\end{align}
	
	In a similar manner we can show (see Appendix~\ref{appendixB}) that the eigenvalues $E_{\textsc{ab},i}$ of $\hat{\rho}_\textsc{ab}$ are:
	\begin{align}
	    E_{\textsc{ab},1}&=1-\mathcal{L}_\textsc{aa}- \mathcal{L}_\textsc{bb}+ \mathcal{O}(\lambda_\nu^3),
	    \\
	    E_{\textsc{ab},2}&=0+\mathcal{O}(\lambda_\nu^3),
	    \\
	    E_{\textsc{ab},3}&=\frac{1}{2}\left(\mathcal{L}_\textsc{aa}+
		\mathcal{L}_\textsc{bb}+
		\sqrt{(\mathcal{L}_\textsc{aa}-\mathcal{L}_\textsc{bb})^2+
		4|\mathcal{L}_\textsc{ab}|^2}\right)
		\notag\\
		&\phantom{=}
		+\mathcal{O}(\lambda_\nu^3),
		\\
		E_{\textsc{ab},4}&=\frac{1}{2}\left(\mathcal{L}_\textsc{aa}+
		\mathcal{L}_\textsc{bb}-
		\sqrt{(\mathcal{L}_\textsc{aa}-\mathcal{L}_\textsc{bb})^2+
		4|\mathcal{L}_\textsc{ab}|^2}\right)
		\notag \\
		&\phantom{=}+\mathcal{O}(\lambda_\nu^3).
	\end{align}	
Note that these eigenvalues are all non-negative to leading order, as shown in Appendix~\ref{appendixB}.	We see that, to $\mathcal{O}(\lambda_\nu^2)$, the eigenvalues $E_{\textsc{ab},i}$ of $\hat{\rho}_\textsc{ab}$ and the eigenvalues $E_{\textsc{ab},i}^{{\text{\textbf{t}}}_\textsc{b}}$ of $\hat{\rho}_\textsc{ab}^{{\text{\textbf{t}}}_\textsc{b}}$ are fully determined by the vacuum terms $\mathcal{L}_\textsc{aa}$, $\mathcal{L}_\textsc{bb}$, $\mathcal{L}_\textsc{ab}$, and $\mathcal{M}$, and are thus independent of the coherent amplitude of the field. This completes the proof of Theorem 2.

   In the context of entanglement harvesting, we are interested in the eigenvalues $E_{\textsc{ab},i}^{{\text{\textbf{t}}}_\textsc{b}}$, since, as discussed above, they determine the negativity of the two-detector system following its interaction with the field. We have shown that the $E_{\textsc{ab},i}^{{\text{\textbf{t}}}_\textsc{b}}$ are, to $\mathcal{O}(\lambda_\nu^2)$, equal to the eigenvalues of the two-detector density matrix calculated in~\cite{Pozas2015}, where the authors considered a pair of detectors interacting with the vacuum state of the field. Hence the results from~\cite{Pozas2015} regarding the dependence of vacuum entanglement harvesting on detector parameters and the spacetime dimensionality apply to entanglement harvesting from any general coherent state.
   
There is a remarkable highlight of this result: one would perhaps have expected that `shining a laser' on a couple of detectors would have increased the local noise (indeed, as opposed to the vacuum case, there appear order $\lambda_\nu$ terms in the detectors excitation probabilities, which are intrinsically local and leading-order), and as such, decrease the ability of the detectors to harvest entanglement \cite{Reznik2005,Pozas2015}. However, we find that the added leading order local noise in the case of coherent states has no impact on the detector pair's ability to harvest entanglement since its effect gets cancelled exactly by extra correlation terms due to the coherent nature of the field state. Mathematically, we see how first order local noise terms only appear in the negativity with higher powers of $\lambda_\nu$, and in a way that cancels exactly with new second order terms, so as to leave the amount of harvestable entanglement invariant with respect to changes in the coherent amplitude of the field.

	\section{Conclusions}
	\label{sec:conclusions}
	
	We studied the effects of interacting with a general coherent state of a scalar field on the dynamics of one and two Unruh-DeWitt particle detectors.
	
	For the case of a single two-level detector initialized to its ground state, unsurprisingly, the detector is more likely to make a transition into its excited state if the field is in an non-trivial coherent state rather than the vacuum. Remarkably however, we found that the eigenvalues of the time-evolved detector's density matrix following its interaction with the field are independent of the  coherent amplitude of the field state, at least to second order in the detector-field coupling strength $\lambda_\textsc{a}$. Our result shows, therefore, that any property of the detector that is determined by the eigenvalues of its density matrix is the same regardless of whether the detector interacts with a general coherent field state or the vacuum. One such property is the von Neumann entropy, implying, for example, that if we have an ensemble of detectors in their ground states, and we let each one interact with a different coherent state of the field, following the interaction all of the detectors' states will be equally mixed.
	
	Interestingly, we obtained an analogous result in the case where we have a pair of detectors, initialized to their ground states, interacting with the field. Namely we found that, to second order in the detectors' coupling strengths to the field, the final state density matrix of these detectors has a spectrum that is independent of the coherent amplitude of the field. We also found this to be true for the eigenvalues of the partially transposed density matrix. A particular consequence of the latter result is that the negativity of the two-detector system (an entanglement measure often investigated in the context of entanglement harvesting \cite{Reznik2003}) does not depend on which coherent state the field was in. This is particularly interesting since a non-vanishing coherent amplitude of the field introduces inherently local leading order corrections to the two-detector density matrix, and one may have in principle expected this to introduce local noise, which is detrimental to entanglement harvesting \cite{Reznik2005,Pozas2015}. We have seen that this is not the case for coherent states.
	
	\section*{Acknowledgments}
	
    The work of P. S. and E. M.-M. is supported by the Natural Sciences and Engineering Research Council of Canada through the USRA and Discovery programs. E. M.-M. also gratefully acknowledges the funding of his Ontario Early Research Award.

	\appendix
	
	\section{Explicit computation of \texorpdfstring{$\bar{L}_\nu$}{}, \texorpdfstring{$\mathcal{\bar{M}}$}{} and \texorpdfstring{$\mathcal{\bar{L}}_{\mu\nu}$}{}}
	\label{appendixA}
	
    We define $\bar{L}_\textsc{a}\coloneqq (\hat{\rho}_{\textsc{ab}}^{(1)})_{(3,1)}$ to be the $(3,1)$ component of the matrix representation~\eqref{eq:rho_ab_coh} of $\hat{\rho}_{\textsc{ab}}$ in the basis~\eqref{eq:basis}. Hence, using~\eqref{eq:rho_ab_1},
    \begin{align}
        \bar{L}_\textsc{a}&=
        \ii\!\!\!\!\sum_{\nu\in\{\text{A,B}\}}\!\!\!\!\lambda_\nu
		\int_{-\infty}^{\infty}\!\!\!\!\!\dif t\,\chi_\nu(t)
		([\hat{\rho}_{\textsc{ab},0},\hat{\mu}_\nu(t)])_{(3,1)}V(\bm{x}_\nu,t)
		\notag\\
		&=
		\label{eq:L_A_app}
		-\ii\lambda_\textsc{a}
		\int_{-\infty}^{\infty}\!\!\!\!\!\dif t\,\chi_\textsc{a}(t)
		e^{\ii\Omega_\textsc{a}t}V(\bm{x}_\nu,t),
    \end{align}
    where we have evaluated $([\hat{\rho}_{\textsc{ab},0},\hat{\mu}_\nu(t)])_{(3,1)}$ by using the expressions for $\hat{\mu}_\nu(t)$ in~\eqref{eq:monopole} and the expression for $\hat{\rho}_{\textsc{ab},0}$ in~\eqref{eq:rho_ab_0}. Similarly we define $\bar{L}_\textsc{b}\coloneqq (\hat{\rho}_{\textsc{ab}}^{(1)})_{(2,1)}$, which gives
    \begin{align}
        \bar{L}_\textsc{b}&=
        \ii\!\!\!\!\sum_{\nu\in\{\text{A,B}\}}\!\!\!\!\lambda_\nu
		\int_{-\infty}^{\infty}\!\!\!\!\!\dif t\chi_\nu(t)
		([\hat{\rho}_{\textsc{ab},0},\hat{\mu}_\nu(t)])_{(2,1)}V(\bm{x}_\nu,t)
		\notag\\
		&=
		\label{eq:L_B_app}
		-\ii\lambda_\textsc{b}
		\int_{-\infty}^{\infty}\!\!\!\!\!\dif t\chi_\textsc{b}(t)
		e^{\ii\Omega_\textsc{b}t}V(\bm{x}_\nu,t).
    \end{align}
    
    Next we define $\mathcal{M}+\bar{\mathcal{M}}$ to be the $(4,1)$ component of~\eqref{eq:rho_ab_coh}. Using~\eqref{eq:rho_ab_2} this gives:
    \begin{align}
        \label{eq:M+M_bar}
        \mathcal{M}&+\bar{\mathcal{M}}
        \coloneqq
        \!\!\!\!\sum_{\nu,\eta\in\{\text{A,B}\}}\!\!\!\!
		\lambda_\nu\lambda_\eta
		\Bigg[\int_{-\infty}^{\infty}\!\!\!\!\!\dif t
			\int_{-\infty}^{\infty}\!\!\!\!\!\dif t'
			\chi_\nu(t')\chi_\eta(t) \notag\\
			&\phantom{{}=
				\!\!\!\!\sum_{\nu,\eta\in\{\textsc{a,b}\}}\!\!\!\!
				\lambda_\nu}
			\times
			(\hat{\mu}_\nu(t')
	    	\hat{\rho}_{\textsc{ab},0}
	    	\hat{\mu}_\eta(t))_{(4,1)}
			W(\bm{x}_\eta,t,\bm{x}_\nu,t')\notag\\
			&\phantom{{}=\,\,}
			-\int_{-\infty}^{\infty}\!\!\!\!\!\dif t
			\int_{-\infty}^{t}\!\!\!\!\!\dif t'
			\chi_\nu(t)\chi_\eta(t') \notag\\
			&\phantom{{}=
				\!\!\!\!\sum_{\nu,\eta\in\{\textsc{a,b}\}}\!\!\!\!
				\lambda_\nu}
			\times
			(\hat{\mu}_\nu(t)
			\hat{\mu}_\eta(t')
			\hat{\rho}_{\textsc{ab},0})_{(4,1)}
			W(\bm{x}_\nu,t,\bm{x}_\eta,t')\notag\\
			&\phantom{{}=\,\,}
			-\int_{-\infty}^{\infty}\!\!\!\!\!\dif t
			\int_{-\infty}^{t}\!\!\!\!\!\dif t'
			\chi_\nu(t)\chi_\eta(t') \notag\\
			&\phantom{{}=
				\!\!\!\!\sum_{\nu,\eta\in\{\textsc{a,b}\}}\!\!\!\!
				\lambda_\nu}
			\times
			(\hat{\rho}_{\textsc{ab},0}\hat{\mu}_\eta(t')\hat{\mu}_\nu(t))_{(4,1)}
			W(\bm{x}_\eta,t',\bm{x}_\nu,t)
		\Bigg].
	\end{align}
	Recall from~\eqref{eq:W_coh} that for coherent field states $W(\bm{x}_\nu,t,\bm{x}_\eta,t')$ can be written as 
	\begin{equation}
	    W(\bm{x}_\nu,t,\bm{x}_\eta,t')
	    =
		V(\bm{x}_\nu,t) V(\bm{x}_\eta,t')+
		W_\text{vac}(\bm{x}_\nu,t,\bm{x}_\eta,t'),
	\end{equation}
	where $V(\bm{x}_\nu,t)V(\bm{x}_\eta,t')$ is an $\alpha(\bm k)$-dependent term and $W_\text{vac}(\bm{x}_\nu,t,\bm{x}_\eta,t')$ is a vacuum term. We define $\bar{\mathcal{M}}$ to be the $\alpha(\bm k)$-dependent component of~\eqref{eq:M+M_bar}, and $\mathcal{M}$ to be the vacuum component of~\eqref{eq:M+M_bar}:
	\begin{align}
	\label{eq:M_bar_app}
		\bar{\mathcal{M}}
		&\coloneqq
		-\lambda_\textsc{a}
		\lambda_\textsc{b}
		\int_{-\infty}^{\infty}\!\!\!\!\!\dif t
		\int_{-\infty}^{t}\!\!\!\!\!
		\dif t'
		\notag\\
		&\phantom{\coloneqq}
		\times
		\big[
			\chi_\textsc{a}(t)
			\chi_\textsc{b}(t')
			e^{\ii\Omega_\textsc{a}t}
			e^{\ii\Omega_\textsc{b}t'}
	    	V(\bm{x}_\textsc{a},t)
	    	V(\bm{x}_\textsc{b},t')
        \notag\\
        &\phantom{\coloneqq}+
            \chi_\textsc{b}(t)
			\chi_\textsc{a}(t')
			e^{\ii\Omega_\textsc{b}t}
			e^{\ii\Omega_\textsc{a}t'}
			V(\bm{x}_\textsc{b},t)
	    	V(\bm{x}_\textsc{a},t')
		\big],
		\\
		\mathcal{M}
		&\coloneqq
		-\lambda_\textsc{a}
		\lambda_\textsc{b}
		\int_{-\infty}^{\infty}\!\!\!\!\!\dif t
		\int_{-\infty}^{t}\!\!\!\!\!
		\dif t'
		\notag\\
		&\phantom{\coloneqq}
		\times
		\big[
			\chi_\textsc{a}(t)
			\chi_\textsc{b}(t')
			e^{\ii\Omega_\textsc{a}t}
			e^{\ii\Omega_\textsc{b}t'}
	    	W_\text{vac} (\bm{x}_\textsc{a},t,\bm{x}_\textsc{b},t')
        \notag\\
        &\phantom{\coloneqq}+
            \chi_\textsc{b}(t)
			\chi_\textsc{a}(t')
			e^{\ii\Omega_\textsc{b}t}
			e^{\ii\Omega_\textsc{a}t'}
			W_\text{vac} (\bm{x}_\textsc{b},t,\bm{x}_\textsc{a},t')
		\big].
    \end{align}
    The vacuum term $\mathcal{M}$ has been calculated in~\cite{Pozas2015}. Meanwhile we can simplify the expression for $\bar{\mathcal{M}}$ in~\eqref{eq:M_bar_app} by relabeling the integration variables in the second term, thereby rewriting the expression as an integral over the entire $t$-$t'$ plane:
    \begin{align}
		\bar{\mathcal{M}}&=
		-\lambda_\textsc{a}
		\lambda_\textsc{b}
		\int_{-\infty}^{\infty}\!\!\!\!\!\dif t
		\int_{-\infty}^{\infty}\!\!\!\!\!\dif t'
		\notag\\
		&\phantom{=-}
		\times
			\chi_\textsc{a}(t)
			\chi_\textsc{b}(t')
			e^{\ii\Omega_\textsc{a}t}
			e^{\ii\Omega_\textsc{b}t'}
	    	V(\bm{x}_\textsc{a},t)
	    	V(\bm{x}_\textsc{b},t').
    \end{align}
	We can rewrite this using~\eqref{eq:L_A_app} and~\eqref{eq:L_B_app} as
	\begin{equation}
	    \bar{\mathcal{M}}=
	    \bar{L}_\textsc{a}
	    \bar{L}_\textsc{b}.
	\end{equation}
	
	Next, we define $\mathcal{L}_\textsc{aa}+\bar{\mathcal{L}}_\textsc{aa}$ to be the (3,3) component of~\eqref{eq:rho_ab_coh}, which comes from the first summand in~\eqref{eq:rho_ab_2}. We define $\bar{\mathcal{L}}_\textsc{aa}$ to be the $\alpha(\bm k)$-dependent term in this component. This gives
	\begin{align}
	\label{eq:L_aa_bar}
		\bar{\mathcal{L}}_\textsc{aa}
		&=
		\lambda_\textsc{a}^2
		\int_{-\infty}^{\infty}\!\!\!\!\!\dif t
		\int_{-\infty}^{\infty}\!\!\!\!\!\dif t'
		\notag\\
		&\phantom{=-}
		\times
			\chi_\textsc{a}(t)
			\chi_\textsc{a}(t')
			e^{-\ii\Omega_\textsc{a}t}
			e^{\ii\Omega_\textsc{a}t'}
	    	V(\bm{x}_\textsc{a},t)
	    	V(\bm{x}_\textsc{a},t')
	    \notag\\
	    &=
	    \bar{L}_\textsc{a}
	    \bar{L}_\textsc{a}^*.
    \end{align}
    Similarly, we define $\bar{\mathcal{L}}_\textsc{bb}$ to be the $\alpha(\bm k)$-dependent term in $(\hat{\rho}_{\textsc{ab}}^{(2)}) _{(2,2)}$, and we obtain
    \begin{equation}
    \label{eq:L_bb_bar}
        \mathcal{L}_\textsc{bb}=
        \bar{L}_\textsc{b}
	    \bar{L}_\textsc{b}^*.
    \end{equation}
    
    Finally we define $\mathcal{L}_\textsc{ab}+\bar{\mathcal{L}}_\textsc{ab}$ to be the (3,2) component of~\eqref{eq:rho_ab_coh}, and we define $\bar{\mathcal{L}}_\textsc{ab}$ to be the term in this component that depends on $\alpha(\bm k)$. This gives:
    \begin{align}
    \label{eq:L_ab_bar}
		\bar{\mathcal{L}}_\textsc{ab}
		&=
		\lambda_\textsc{a}
		\lambda_\textsc{b}
		\int_{-\infty}^{\infty}\!\!\!\!\!\dif t
		\int_{-\infty}^{\infty}\!\!\!\!\!\dif t'
		\notag\\
		&\phantom{=-}
		\times
			\chi_\textsc{a}(t)
			\chi_\textsc{b}(t')
			e^{-\ii\Omega_\textsc{b}t}
			e^{\ii\Omega_\textsc{a}t'}
	    	V(\bm{x}_\textsc{b},t)
	    	V(\bm{x}_\textsc{a},t')
	    \notag\\
	    &=
	    \bar{L}_\textsc{a}
	    \bar{L}_\textsc{b}^*.
    \end{align}    
    We can succinctly write expressions~\eqref{eq:L_aa_bar},~\eqref{eq:L_bb_bar} and~\eqref{eq:L_ab_bar} as
	\begin{equation}
	    \bar{\mathcal{L}}_{\mu\nu}=
		\bar{L}_\mu\bar{L}_\nu^*.
	\end{equation}

\section{Computing the spectrum of \texorpdfstring{$\hat{\rho}_\text{AB}$}{}}
\label{appendixB}

The eigenvalues of $\hat{\rho}_\textsc{ab}$ (equation~\eqref{eq:rho_ab_coh}) are the roots of the characteristic polynomial
	\begin{align}
		q(y)&=y^4+[-1 +\mathcal{O}(\lambda_\nu^3)]y^3
		+[C_2'
		+\mathcal{O}(\lambda_\nu^3)]y^2
		\notag\\
		\label{eq:char_poly'}
		&\phantom{=}
		+[C_4'
		+\mathcal{O}(\lambda_\nu^5)]y
		+\mathcal{O}(\lambda_\nu^7),
	\end{align}
	with $C_2'\propto\lambda_\nu^2$ and $C_4'\propto\lambda_\nu^4$ defined as
	\begin{align}
	    \label{eq:C_2'}
        C_2'&\coloneqq
        \mathcal{L}_\textsc{aa}+
        \mathcal{L}_\textsc{bb},
        \\
		C_4'&\coloneqq
		|\mathcal{L}_\textsc{ab}|^2-
		\mathcal{L}_\textsc{aa}
		\mathcal{L}_\textsc{bb}.
	\end{align}
	Notice that the leading order coefficients of each power of $y$ in $q(y)$ are independent of the coherent amplitude $\alpha(\bm k)$ (the $\bar{\mathcal{L}}_{\mu\nu}= \bar{L}_\mu\bar{L}_\nu^*$, $\bar{\mathcal{M}}=\bar{L}_\textsc{a} \bar{L}_\textsc{b}$, and $\bar{L}_\nu$ terms cancel exactly in the calculation of the determinant leading to $q(y)$). The most general form for a root of $q(y)$ is given by
	\begin{equation}
	    Y=Y_0+Y_1+Y_2+
	    \mathcal{O}(\lambda_\nu^3),
	\end{equation}
    where the $Y_i$ term is $\mathcal{O}[(\lambda_\nu)^i]$. We now substitute $Y$ into $q(y)$ and systematically set terms proportional to $(\lambda_\nu)^i$ equal to zero, starting from $i=0$ and moving up to higher orders until we determine each $Y_i$. To that end, equating the terms of $q(Y)$ proportional to $\lambda_\nu^0$ to zero, we obtain $0=Y_0^3(Y_0-1)$. Hence $Y_0=0$ or $Y_0=1$.
    
    \textbf{Case 1}: $Y_0=1$. \\
    Equating the $\lambda_\nu^1$ terms in $q(Y)$ to zero, we obtain $Y_1=0$. The $\lambda_\nu^2$ terms give $Y_2=-C_2'$. Therefore, using the definition~\eqref{eq:C_2'} of $C_2'$, the first root $E_{\textsc{ab},1}$ of the characteristic polynomial $q(y)$ is
    \begin{equation}
        E_{\textsc{ab},1}=1-\mathcal{L}_\textsc{aa}- \mathcal{L}_\textsc{bb}+ \mathcal{O}(\lambda_\nu^3).
    \end{equation}
	
	\textbf{Case 2}: $Y_0=0$. \\
	In this case there are no terms in $q(Y)$ proportional to $\lambda_\nu^1$ or $\lambda_\nu^2$. Equating the terms proportional to $\lambda_\nu^3$ to zero gives $Y_1=0$. There are no terms proportional to $\lambda_\nu^4$ or $\lambda_\nu^5$. Equating the terms proportional to $\lambda_\nu^6$ to zero gives $0=-Y_2(Y_2^2-C_2'Y_2-C_4')$. Hence $Y_2=0$ or \mbox{$Y_2=(C_2'\pm\sqrt{C_2'^2+4C_4'})/2$}. Therefore the remaining roots of $q(y)$ are
	\begin{align}
	    E_{\textsc{ab},2}&=0+\mathcal{O}(\lambda_\nu^3),
	    \\
	    E_{\textsc{ab},3}&=\frac{1}{2}\left(\mathcal{L}_\textsc{aa}+
		\mathcal{L}_\textsc{bb}+
		\sqrt{(\mathcal{L}_\textsc{aa}-\mathcal{L}_\textsc{bb})^2+
		4|\mathcal{L}_\textsc{ab}|^2}\right)
		\notag\\
		&\phantom{=}
		+\mathcal{O}(\lambda_\nu^3),
		\\
		\label{eq:E_ab4}
		E_{\textsc{ab},4}&=\frac{1}{2}\left(\mathcal{L}_\textsc{aa}+
		\mathcal{L}_\textsc{bb}-
		\sqrt{(\mathcal{L}_\textsc{aa}-\mathcal{L}_\textsc{bb})^2+
		4|\mathcal{L}_\textsc{ab}|^2}\right)
		\notag \\
		&\phantom{=}+\mathcal{O}(\lambda_\nu^3).
	\end{align}

Since the eigenvalues of $\hat{\rho}_\textsc{ab}$ depend only on the vacuum terms $\mathcal{L}_\textsc{aa}$, $\mathcal{L}_\textsc{bb}$, and $\mathcal{L}_\textsc{ab}$, this concludes the proof that these eigenvalues are independent of the coherent amplitude $\alpha(\bm k)$.

As a consistency check, we can ensure that the eigenvalues of the positive semi-definite density operator $\hat{\rho}_\textsc{ab}$ are non-negative. $E_{\textsc{ab},1}$ is non-negative by assumption, since we assume that the coupling constants $\lambda_\textsc{a}$ and $\lambda_\textsc{b}$, which appear in $\mathcal{L}_\textsc{aa}$ and  $\mathcal{L}_\textsc{bb}$, are small (if $E_{\textsc{ab},1}$ were negative, $\lambda_\textsc{a}$ and $\lambda_\textsc{b}$ would not be small enough for us to work in the perturbative regime as we have done). Also, $E_{\textsc{ab},2}$ is zero, and hence non-negative, to $\mathcal{O}(\lambda_\nu^2)$.

To show that $E_{\textsc{ab},3}$ and $E_{\textsc{ab},4}$ are non-negative, we first recall the definition of $\mathcal{L}_{\mu\nu}$ from~\eqref{eq:L_mu_nu}.
\begin{equation}
    \label{eq:L_mu_nu2}
    \mathcal{L}_{\mu\nu}
	=		\int\d[n]{\bm{k}}L_\mu(\bm{k})L_\nu(\bm{k})^*.
\end{equation}
Clearly $\mathcal{L}_\textsc{aa}$ and $\mathcal{L}_\textsc{bb}$ are non-negative, and therefore so is $E_{\textsc{ab},3}$. Proving that $E_{\textsc{ab},4}$ is non-negative is more involved. We start with a lemma.

\textbf{Lemma}: Let $(a_i)_{i=1}^n$ and $(b_i)_{i=1}^n$ be two sequences of non-negative real numbers. Then
\begin{equation}
    \label{eq:lemma}
    \sum_{i=1}^n a_i^2 
    \sum_{j=1}^n b_j^2 
    \ge
    \left(\sum_{i=1}^n a_i b_i \right)^2.
\end{equation}
\textit{Proof}: We first evaluate the left-hand side of~\eqref{eq:lemma}.
\begin{align}
    \label{eq:sum1}
    \sum_{i=1}^n a_i^2
    \sum_{j=1}^n b_j^2
    &=
    \sum_{1\le i,j\le n} a_i^2 b_j^2
    \notag\\
    &=    
    \sum_{i=1}^n a_i^2 b_i^2+\!\!\!
    \sum_{1\le i<j \le n} 
    \!\!\!(a_i^2 b_j^2 +a_j^2 b_i^2).
\end{align}
Meanwhile the right-hand side of~\eqref{eq:lemma} evaluates to
\begin{align}
    \label{eq:sum2}
    \left(\sum_{i=1}^n a_i b_i \right)^2
    &=
    \sum_{1\le i,j\le n} a_i b_i a_j b_j
    \notag\\
    &=    
    \sum_{i=1}^n a_i^2 b_i^2+\!\!\!
    \sum_{1\le i<j \le n} 
    \!\!\!2 a_i b_i a_j b_j.
\end{align}
Using the fact that
\begin{equation}
    0
    \le 
    (a_i b_j- a_j b_i)^2
    =
    a_i^2 b_j^2 +a_j^2 b_i^2 - 2a_ib_ia_jb_j,
\end{equation}
we obtain that $a_i^2 b_j^2 +a_j^2 b_i^2 \ge 2a_ib_ia_jb_j$, which, together with~\eqref{eq:sum1} and~\eqref{eq:sum2} proves the lemma.

\textbf{Corollary}: Let $f_\textsc{a}(\bm k)$ and $f_\textsc{b}(\bm k)$ be real-valued, non-negative, continuous, functions of $\bm k$. Then, taking the continuum limit of the sums in the lemma, one obtains that
\begin{equation}
    \label{eq:corollary}
    \int\d[n]{\bm k} f_\textsc{a}(\bm k)^2
    \int\d[n]{\bm k'} f_\textsc{b}(\bm k')^2
    \ge 
    \left(
    \int\d[n]{\bm k} f_\textsc{a}(\bm k) f_\textsc{b}(\bm k)
    \right)^2.
\end{equation}

Now, define $L_\nu(\bm k)\coloneqq f_\nu(\bm k) e^{\ii\theta_\nu(\bm k)}$ for $\nu\in\{\text{A,B}\}$, with $f_\nu(\bm k)$ being real-valued, continuous, non-negative functions of $\bm k$. Then, using the definition~\eqref{eq:L_mu_nu2} we can rewrite $\mathcal{L}_\textsc{aa} \mathcal{L}_\textsc{bb}$ as
\begin{equation}
    \label{eq:LaaLbb}
    \mathcal{L}_\textsc{aa} \mathcal{L}_\textsc{bb}
    =
    \int\d[n]{\bm k} f_\textsc{a}(\bm k)^2
    \int\d[n]{\bm k'} f_\textsc{b}(\bm k')^2.
\end{equation}
On the other hand, we can write $|\mathcal{L}_\textsc{ab}|^2$ as
\begin{align}
    \label{eq:Lab^2}
    |\mathcal{L}_\textsc{ab}|^2
    &=
    \left|\int\d[n]{\bm k} f_\textsc{a}(\bm k) f_\textsc{b}(\bm k) e^{\ii(\theta_\textsc{a}(\bm k)-\theta_\textsc{b}(\bm k))}\right|
    \notag\\
    &\phantom{=}
    \times 
    \left|\int\d[n]{\bm k'} f_\textsc{a}(\bm k') f_\textsc{b}(\bm k') e^{-\ii(\theta_\textsc{a}(\bm k')-\theta_\textsc{b}(\bm k'))}\right|.
\end{align}
Note that for a real-valued, non-negative function $f(\bm k)$ of $\bm k$, and a real-valued function $\theta(\bm k)$ of $\bm k$, it is easily shown that
\begin{equation}
    \left|\int\d[n]{\bm k} f(\bm k) e^{\ii \theta(\bm k)}\right|
    \le
    \int\d[n]{\bm k} f(\bm k).
\end{equation}
Hence, from~\eqref{eq:Lab^2}, we obtain
\begin{equation}
    \label{eq:Lab^2_2}
    |\mathcal{L}_\textsc{ab}|^2
    \le
    \left(
    \int\d[n]{\bm k} f_\textsc{a}(\bm k) f_\textsc{b}(\bm k)
    \right)^2.
\end{equation}
Then, using~\eqref{eq:LaaLbb},~\eqref{eq:Lab^2_2}, and the corollary~\eqref{eq:corollary}, one obtains
\begin{equation}
    \label{eq:inequality}
    \mathcal{L}_\textsc{aa} \mathcal{L}_\textsc{bb}
    \ge 
    |\mathcal{L}_\textsc{ab}|^2.
\end{equation}
Using this, we can obtain a bound on the radicand in expression~\eqref{eq:E_ab4} for $E_{\textsc{ab},4}$,
\begin{align}
    \label{eq:bound}
    &\phantom{=}\,\,\,
    (\mathcal{L}_\textsc{aa}-\mathcal{L}_\textsc{bb})^2 
    +
    4|\mathcal{L}_\textsc{ab}|^2
    \notag\\
    &=
    (\mathcal{L}_\textsc{aa}+\mathcal{L}_\textsc{bb})^2
    - 4 \mathcal{L}_\textsc{aa} \mathcal{L}_\textsc{bb}
    + 4|\mathcal{L}_\textsc{ab}|^2
    \notag\\
    &\le
    (\mathcal{L}_\textsc{aa}+\mathcal{L}_\textsc{bb})^2,
\end{align}
which proves that $E_{\textsc{ab},4}\ge 0$. Hence all of the eigenvalues $E_{\textsc{ab},i}$ of the density matrix $\hat{\rho}_\textsc{ab}$ are non-negative.

	\bibliography{references_coh}
	\bibliographystyle{apsrev4-1}

\end{document}